\documentclass[preprint]{article}
\usepackage{authblk}
\usepackage{graphicx}
\usepackage{dcolumn}
\usepackage{bm}
\usepackage{amsmath}
\usepackage{amssymb}
\usepackage{multirow}
\usepackage{caption}
\usepackage{subcaption}
\usepackage{epstopdf}
\usepackage{hyperref}
\usepackage{cite}
\begin{document}
\title{\bf Baryogenesis constraints and parameter bounds in $f(T,T_{G})$ modified gravity}
\author[]{Amit Samaddar\thanks{samaddaramit4@gmail.com}}
\author[]{S. Surendra Singh\thanks{ssuren.mu@gmail.com}}
\affil[]{Department of Mathematics, National Institute of Technology Manipur, Imphal-795004,India.}

\maketitle

\begin{abstract}
We investigate the generation of the observed baryon asymmetry of the Universe within the framework of $f(T,T_{G})$ gravity, where $T$ is the torsion scalar and $T_{G}$ denotes its teleparallel Gauss--Bonnet counterpart. Two illustrative models, $f(T,T_{G})=\alpha T+\beta \sqrt{T_{G}}$ and $f(T,T_{G})=-T+\delta\, T_{G}\ln(T_{G})$, are examined in a power-law background $a(t)=a_{0} t^{m}$. For both models, we derive analytic expressions for the baryon-to-entropy ratio $\eta_{B}/s$ using the standard and generalized baryogenesis formalisms, adopting high-energy decoupling conditions with $g_{b}=1$, $g_{s}=106$, $T_{D}=2\times10^{16}\,\mathrm{GeV}$, and $M_{\star}=2\times10^{12}\,\mathrm{GeV}$. Consistency of the cosmological dynamics requires $m>1$, and the observed value $\eta_{B}/s \simeq 9.42\times10^{-11}$ is obtained for constrained intervals of the parameters $\alpha$, $\beta$, $\delta$, and $m$. Numerical results confirm that both models reproduce the measured baryon asymmetry without invoking extra fields or exotic matter sources. These findings indicate that teleparallel gravity with a Gauss--Bonnet torsion term provides a natural and viable mechanism for baryogenesis, offering a compelling alternative to curvature-based descriptions of the early Universe.
\end{abstract}


\section{Introduction}\label{sec1}
\hspace{0.5cm} The question of how the Universe developed its excess of matter over antimatter continues to be a core problem driving research in cosmology and particle physics. In the conventional cosmological framework, the Universe began in a highly energetic and compact state following the Big Bang, with particle interactions occurring swiftly and in a symmetric manner \cite{Wilczek1980}. Given these circumstances, the creation of baryons and antibaryons would be symmetric and they would destroy one another as the Universe dilated and its temperature dropped. Astronomical data show that the current Universe is overwhelmingly composed of matter, while any initial antimatter has effectively disappeared. The observed discrepancy between matter and antimatter is encapsulated in the baryon-to-entropy ratio, $\frac{\eta_B}{s}\approx 9 \times 10^{-11}$, determined from high-precision Big Bang Nucleosynthesis (BBN) analyses and Cosmic Microwave Background observations \cite{Bennett2003,Burles2001}. Explaining this tiny but critical imbalance requires mechanisms beyond the basic predictions of the standard model. The theoretical requirements for generating a baryon asymmetry were formulated by Sakharov \cite{Sakharov1967}, who argued that three criteria are necessary for any successful baryogenesis scenario: (i) interactions that violate baryon number conservation, (ii) violation of C and CP symmetries so that baryons and antibaryons are produced at different rates and (iii) a departure from thermal equilibrium to prevent reverse reactions from continuously erasing any asymmetry. They provide the fundamental basis upon which most conventional baryogenesis models are built.

Numerous theoretical frameworks have been established in particle physics to fulfill these requirements. Electroweak baryogenesis explains the generation of baryon asymmetry at the electroweak epoch, requiring a substantial CP-violating effect \cite{Trodden1999}. In models with scalar fields or supersymmetric extensions, new dynamical degrees of freedom naturally lead to the production of a baryon asymmetry, as seen in the Affleck–Dine approach \cite{Stewart1996}. Leptogenesis, whether thermal or non-thermal, establishes a connection between lepton-number generation and the resulting baryon asymmetry through sphaleron-mediated processes \cite{Dolgov1981}. While conceptually promising, the majority of these scenarios invoke physics beyond current accelerator limits or demand CP-violating strengths larger than those observed in experiments. Davoudiasl and colleagues \cite{Davou2004} made a significant contribution by extending spontaneous baryogenesis to apply within a thermally equilibrated, expanding cosmological background. Their approach is based on the idea that the curvature of spacetime changes over cosmic time, and this dynamical evolution can generate CP violation. In this framework, the baryon current $J_\mu$ is coupled to the derivative of the Ricci scalar $R$, producing an effective interaction in the action given by $\frac{1}{M_\star^2}\int \sqrt{-g} \, (\partial_\mu R) J^\mu \, d^4x,$ where $M_\star$ is the characteristic high-energy cutoff and $g$ is the determinant of the metric. As the Universe evolves, this coupling distinguishes baryons from antibaryons, enabling the evolving curvature to induce a net baryon asymmetry without departing from thermal equilibrium. Within a spatially flat FLRW background, the Ricci scalar depends solely on cosmic time, and this interaction produces an energy-level asymmetry between matter and antimatter. Its contribution to the baryon current is given by $\frac{1}{M_\star^2} (\partial_\mu R) J_B^\mu = \frac{\dot{R}}{M_\star^2} \left(\eta_B - \eta_{\bar{B}}\right)$, with $\dot{R}$ representing the temporal derivative of $R$, and $\eta_B$, $\eta_{\bar{B}}$ the respective number densities of baryons and antibaryons. This mechanism induces a CPT-violating shift in the chemical potential for both baryons and antibaryons, described by $\mu_B = - \mu_{\bar{B}} = -\frac{\dot{R}}{M_\star^2}$. At a finite temperature, this chemical potential affects the statistical equilibrium of baryons and antibaryons. The baryon number density at temperature $T$ is then given by $\eta_B = \frac{g_b T^3}{6\pi^2} \left( \frac{\pi^2 \mu_B}{T} + \left( \frac{\mu_B}{T} \right)^3 \right),$ where $g_b$ represents the number of internal degrees of freedom for baryonic species.

General Relativity (GR) has been extremely effective within solar-system tests and astrophysical contexts; nonetheless, observations like the Universe’s accelerated expansion, the presence of dark energy and the challenges posed by dark matter imply that GR might require extensions at cosmological scales \cite{Clifton12}. These observational discrepancies have motivated a significant research effort into modified theories of gravity, frameworks designed to generalize or reconceptualize GR via the inclusion of supplementary geometric or dynamical fields. The theoretical landscape of modified gravity is often expanded by building upon the Einstein-Hilbert action. Theorists introduce additional components based on curvature, torsion, or non-metricity, thereby creating a more complex and powerful geometric description. This enhanced framework is designed to account for cosmological phenomena that lie beyond the scope of standard GR. This pursuit has led to the development of various geometric formulations of gravity that depart from the standard paradigm. Distinct from the curvature-based description of GR, Teleparallel Gravity (TEGR) constitutes a geometric framework in which the gravitational interaction is mediated by torsion, fundamentally characterized by the torsion scalar $T$ \cite{CM61,K79}. Completing the geometric trinity, Symmetric Teleparallel Gravity (STEGR) is a metric-affine theory that frames gravitational interactions in terms of non-metricity, fundamentally described by the non-metricity scalar $Q$ \cite{J18}. The standard procedure of generalizing the Einstein--Hilbert action, exemplified by $f(R)$ gravity where the Ricci scalar $R$ is replaced by a function $f(R)$, can be directly transposed to these alternative geometric descriptions. Consequently, a significant research effort has been dedicated to explore the resulting theories, including $f(T)$, $f(Q)$, $f(R,T)$, $f(R,L_m)$, $f(T,\tau)$, $f(T,\phi)$, $f(T,B)$, $f(Q,T)$ and $f(Q,C)$ gravity \cite{Amit24,Singh23,N21,T11,T08,T10,T14,Amit2,L23,Amit23,S23,AS25,ASAS2025}.

Modified gravity approaches have enriched our understanding not only of the Universe’s accelerated expansion but also of key early-Universe mechanisms, including gravitational baryogenesis. Researchers have considered a variety of gravitational frameworks to study the origin of the baryon–to–entropy ratio $\frac{\eta_B}{s}$. Such studies cover a broad spectrum of gravitational theories, for instance, Randall–Sundrum and Gauss–Bonnet braneworld cosmologies \cite{Bento2005}, modified $f(R)$ models \cite{Lambiase2006,Ramos2017}, Type-IV finite-time singular scenarios \cite{VK16}, various $f(T)$ constructions \cite{VKK16}, loop quantum cosmology together with Gauss–Bonnet baryogenesis \cite{SDOV16}, non-minimal derivative coupling frameworks \cite{Good2023}, $f(R,T)$ gravity \cite{Baffou2019,Sahoo2020} and non-metricity-based models such as $f(Q,T)$ with baryon–current interaction through $\partial_\mu Q$ \cite{SPK2020}. The investigation has been extended to more complex theories such as $f(T,T_G)$ and $f(T,B)$, where $T_G$ is the teleparallel counterpart of the Gauss-Bonnet invariant and $B$ is the boundary term linking torsion and Ricci scalars. For power-law scale factor cosmologies, these frameworks yield consistent predictions for the baryon-to-entropy ratio $\frac{\eta_B}{s}$ \cite{Azhar2020}. Research has also extended to alternative formulations such as $f(G,T)$ gravity, $f(R,G)$ theories \cite{Azhar2021}, $f(R,A)$ models that use the trace of the anti-curvature tensor \cite{Jawad2022}, $f(R,L_m)$ gravity \cite{LV2023,Harko15}, $f(Q,L_{m})$ gravity \cite{AsS2025}, $f(Q,C)$ gravity \cite{ASSS2024,MU24} and modified Ho$\check{r}$ava–Lifshitz scenarios \cite{N20}. Each of these frameworks has produced baryon asymmetry estimates in agreement with observational bounds. In light of these numerous developments, we study gravitational baryogenesis within the context of $f(T,T_G)$ gravity. Two theoretically motivated functions are considered: $f(T,T_G)=f(T,T_{G})=\alpha T+\beta\sqrt{T_{G}}$ and $f(T,T_{G})=-T+\delta\; T_{G}\;\log(T_{G})$. Using these models, we determine the baryon–to–entropy ratio for a spatially flat cosmology characterized by a power-law scale factor $a(t)=a_0 t^{m}$. These models allow us to assess how torsion–Gauss–Bonnet interactions influence the generation of matter–antimatter asymmetry in the early Universe while remaining consistent with the gravitational baryogenesis framework.

The remainder of this paper is organized as follows. In section \ref{sec2}, we summarize the field equations of $f(T,T_{G})$ gravity and introduce the two specific functional models that will be investigated in this study. Section \ref{sec3} presents the theoretical framework for gravitational baryogenesis, including the standard and generalized formulations of the baryon-to-entropy ratio. In this section, we apply the baryogenesis mechanism to the chosen $f(T,T_{G})$ models and analyze the resulting predictions within a power-law cosmological background. Section \ref{sec4} provides concluding remarks and discusses the implications of our results.
\section{Mathematical preliminaries of $f(T,T_{G})$ teleparallel gravity}\label{sec2}
\hspace{0.5cm} This section provides a short overview of the fundamental structure of the $f(T,T_{G})$ modified teleparallel gravity theory, following the conventional approaches discussed in earlier works. Throughout our analysis, Greek indices denote components in the space–time manifold, while Latin indices are used for quantities defined in the tangent (Lorentz) frame. The geometric structure of the theory is determined by the vierbein field $e_{\xi}(x^{\nu})$, which establishes an orthonormal basis in the tangent space. Its components satisfy $e^{\xi}=e^{\xi}{}_{\nu}\,dx^{\nu}$, with the corresponding dual vectors written as $e_{\xi}=e_{\xi}{}^{\nu}\,dx_{\nu}$. In the teleparallel description, the Weitzenb$\ddot{o}$ck connection is adopted as the fundamental affine connection governing parallel transport and its coordinate components are defined as \cite{Kofinas2014}
\begin{equation}\label{1}
w^{\mu}{}_{\nu\sigma}=e_{\xi}{}^{\mu}e^{\xi}{}_{\nu,\sigma}.
\end{equation}
The Weitzenb$\ddot{o}$ck connection transforms non-homogeneously under local Lorentz transformations, and therefore its tangent–space components satisfy $w^{\mu}{}_{\nu\delta}=0$, which ensures that the non-metricity tensor identically vanishes. For an orthonormal vierbein basis, the metric of space–time is constructed via 
\begin{equation}\label{2}
 g_{\nu\sigma}=e^{\xi}{}_{\nu}e^{\gamma}_{\sigma}\eta_{\xi\gamma},
 \end{equation}
with $\eta_{\xi\gamma}=diag[-1,1,1,1]$ denotes the Minkowski metric employed for raising and lowering indices in the tangent space.

After specifying the vierbein fields and the corresponding connection, we proceed to characterize torsion in the teleparallel framework. The torsion tensor is constructed using the relation below.
\begin{equation}\label{3}
T^{\mu}{}_{\nu\sigma} = e_{\xi}{}^{\mu} \left( \partial_{\sigma} e^{\xi}{}_{\nu} - \partial_{\nu} e^{\xi}{}_{\sigma} \right).
\end{equation}
Because the Weitzenb$\ddot{o}$ck connection imposes the teleparallel condition, the associated Riemann curvature tensor vanishes identically. The deviation between the Weitzenb$\ddot{o}$ck connection and the Levi–Civita connection is captured by the contorsion tensor, which is defined as
\begin{equation}\label{4}
K^{\nu\sigma}{}_{\varrho}=-\frac{1}{2}\left(T^{\nu\sigma}{}_{\varrho}-T^{\sigma\nu}{}_{\varrho}-T_{\varrho}{}^{\nu\sigma}\right).
\end{equation}

With the geometric framework established, the torsion tensor $T^{\mu}{}_{\nu\sigma}$ emerges as the fundamental variable for gravity in this formalism. A natural step is to build Lorentz invariants from this tensor, with the simplest non-trivial candidates being quadratic contractions. Among these, a particular combination, known as the torsion scalar $T$, is distinguished:
\begin{equation}\label{5}
T=\frac{1}{4}T^{\nu\sigma\mu}T_{\nu\sigma\mu}+\frac{1}{2}T^{\nu\sigma\mu}T_{\mu\sigma\nu}-T_{\sigma}{}^{\sigma\nu}T^{\mu}{}_{\mu\nu}.
\end{equation}
The physical significance of $T$ becomes apparent when it is used as the gravitational Lagrangian. The resulting theory, Teleparallel Equivalent of General Relativity (TEGR), defined by the action
\begin{equation}\label{6}
S_{\text{TEGR}}=-\frac{1}{2\kappa^2} \int e T\; d^4x ,
\end{equation}
is empirically indistinguishable from GR at the level of field equations. This equivalence provides a foundation for constructing modified theories. By promoting the Lagrangian to a generic function $f(T)$, we obtain a broad class of gravitational models \cite{AmS2024,Sarkar2025}:
\begin{equation}\label{7}
S=\frac{1}{2\kappa^2} \int e\; f(T)\; d^4x.
\end{equation}
We emphasize that the $f(T)$ and $f(R)$ theories are structurally distinct, arising from different geometric foundations (torsion versus curvature) and thus generate different field equations upon variation.

Beyond the quadratic torsion scalar, it is possible to construct more sophisticated torsion-based invariants in teleparallel geometry, following an approach analogous to the formation of higher-order curvature invariants from the Riemann tensor in standard geometric formulations. In Ref. [55], a specific quartic torsion invariant was introduced, denoted as $T_{G}$, defined as: 
\begin{eqnarray}\label{8}
T_G &=&\bigg[K^{\kappa}{}_{\theta\pi}K^{\theta\mu}{}_{\varrho}K^{\nu}{}_{\psi\lambda}K^{\psi\sigma}{}_{\iota} - 2 K^{\kappa\mu}{}_{\pi}K^{\nu}{}_{\theta\varrho}K^{\theta}{}_{\psi\lambda}K^{\psi\sigma}{}_{\iota} + 2 K^{\kappa\mu}{}_{\pi}K^{\nu}{}_{\theta\varrho}K^{\theta\sigma}{}_{\psi}K^{\psi}{}_{\lambda\iota}\\\nonumber
&&+ 2 K^{\kappa\mu}{}_{\pi}K^{\nu}{}_{\theta\varrho}K^{\theta\sigma}{}_{\lambda,\iota}\bigg]\delta^{\pi\varrho\lambda\iota}_{\kappa\mu\nu\sigma},
\end{eqnarray}
where $\delta^{\pi\varrho\lambda\iota}_{\kappa\mu\nu\sigma}$ corresponds to a determinant whose entries are the usual Kronecker symbols. This quantity serves as the teleparallel counterpart of the well-known Gauss–Bonnet curvature invariant,
\begin{equation}\label{9}
G=R^{2}-4R_{\nu\sigma}R^{\nu\sigma}+R_{\nu\sigma\kappa\mu}R^{\nu\sigma\kappa\mu},
\end{equation}
Consequently it acts only as a topological surface term in four-dimensional spacetime. Drawing motivation from modified gravity models that incorporate $f(G)$ modifications to the Einstein–Hilbert action, it is natural to introduce $f(T_{G})$ terms within teleparallel formulations \cite{Nojiri2005}.

Combining the above with the standard torsion scalar $T$, a broad class of generalized teleparallel models is defined through the action \cite{Hoh17,LKD23}
\begin{equation}\label{10}
S=\frac{1}{2\kappa^2} \int e\; f(T,T_{G})\; d^4x,
\end{equation}
which is naturally extendable to dimensions $D > 4$. Since $T_G$ contains terms quartic in torsion, this framework encompasses a richer structure than the conventional $f(T)$ models, allowing for more intricate dynamical behaviour and potential new phenomenology.

To derive the cosmological behavior associated with the action in equation (\ref{9}), we work within a spatially flat FLRW background, whose line element takes the form.
\begin{equation}\label{11}
ds^{2}=-dt^{2}+a(t)^{2}(dx^{2}+dy^{2}+dz^{2}),
\end{equation}
where $a(t)$ represents the cosmic scale factor. This spacetime is obtained from a simple diagonal choice of the vierbein,
\begin{equation}\label{12}
e^{\xi}_{\nu}=diag(1,a(t),a(t),a(t)),
\end{equation}
which, via the metric definition, reproduces the FLRW line element. The associated inverse vierbein has the structure
\begin{equation}\label{13}
e_{\xi}^{\nu}=diag\left(1,\frac{1}{a(t)},\frac{1}{a(t)},\frac{1}{a(t)}\right),
\end{equation}
Using this cosmological tetrad in the computation of the torsion scalar and the teleparallel Gauss–Bonnet term leads to
\begin{equation}\label{14}
T=6H^{2}, \hspace{0.5cm} T_{G}=24H^{2}(\dot{H}+H^{2}),
\end{equation}
with $H=\frac{\dot{a}}{a}$ being the Hubble parameter. These relations form the foundation for constructing the cosmological field equations in $f(T,T_{G})$ gravity.

With the background cosmology established, varying the action with respect to the tetrads yields the modified Friedmann equations. In this framework, both $T$ and the teleparallel Gauss–Bonnet term $T_{G}$ shape the cosmic evolution. Performing the variation in a flat FLRW spacetime results in the generalized field equations.
\begin{equation}\label{15}
\kappa^{2}\rho=\frac{f}{2}-6H^{2}f_{T}-\frac{T_{G}f_{T_{G}}}{2}+12H^{3}\dot{f}_{T_{G}},
\end{equation}
\begin{equation}\label{16}
\kappa^{2}p=-\frac{f}{2}+2(\dot{H}+3H^{2})f_{T}+2H\dot{f}_{T}+\frac{T_{G}f_{T_{G}}}{2}-\frac{T_{G}\dot{f}_{T_{G}}}{3H}-4H^{2}\ddot{f}_{T_{G}}.
\end{equation}
In these expressions, overdots denote time derivatives and we have introduced the abbreviations $f_{T}=\partial f/\partial T$ and $f_{T_{G}} =\partial f / \partial T_{G}$. The equations thus obtained describe the Hubble expansion in the context of generalized teleparallel gravity, where both quadratic torsion contributions and the higher-order term $T_{G}$ play a role. For convenience in our subsequent discussion, we adopt natural units with $\kappa^{2}=1$.\\

Having established the modified Friedmann equations in the framework of $f(T,T_{G})$ gravity, we now proceed to investigate specific non-linear models that are capable of generating non-trivial deviations from teleparallel GR. Whereas linear forms merely rescale the effective gravitational strength without altering the dynamics, non-linear constructions introduce genuine higher-order torsion effects, leading to a broader range of cosmological phenomena. Motivated by recent studies showing that square-root and logarithmic dependencies of the Gauss–Bonnet counterpart $T_{G}$ can lead to late-time cosmic acceleration, cure finite-time singularities, and produce viable inflationary behaviors without invoking an explicit dark energy component, we focus on two representative models:
\begin{equation}\label{17}
\text{Model 1:}\hspace{0.5cm} f(T,T_{G})=\alpha T+\beta\sqrt{T_{G}}\;,
\end{equation}
\begin{equation}\label{18}
\text{Model 2:}\hspace{0.5cm} f(T,T_{G})=-T+\delta\; T_{G}\;\log(T_{G})\;.
\end{equation}
In this first scenario, a minimal non-linear extension is introduced by supplementing the teleparallel sector with a square-root function of the Gauss–Bonnet invariant. Such a term is known to drive late-time acceleration, support stable de Sitter solutions and emulate holographic dark energy in an effective fluid picture \cite{Capoz2016}. The second model, featuring a logarithmic coupling, is inspired by modified Gauss–Bonnet gravity analyses that demonstrate the ability of $T_{G}\log(T_{G})$ type terms to regulate the ultraviolet behavior of the field equations, influence early-Universe dynamics and provide a phenomenologically successful transition between decelerated and accelerated eras \cite{GKOF2014}.
\section{Core principles of gravitational baryogenesis}\label{sec3}
\hspace{0.5cm} Multiple observational probes point toward a clear dominance of matter over antimatter in the Universe. Big Bang Nucleosynthesis (BBN) calculations \cite{Bennett2003}, supported by the absence of photons from large-scale matter–antimatter annihilation, already hinted at this imbalance. Complementary evidence from the CMB anisotropy measurements \cite{Burles2001} leads to similar conclusions. Specifically, BBN predicts $\frac{\eta_{B}}{s}=(5.6\pm0.6)\times 10^{-10}$ and the CMB yields $\frac{\eta_{B}}{s}=(6.19\pm0.14)\times 10^{-10}$. Further elaborations on the baryon asymmetry problem can be found in \cite{SHS2006,Mohanty2006}. More recent observational estimates suggest a present-day value of $\frac{\eta_{B}}{s}\simeq 9.42\times 10^{-11}$, as reported in \cite{Ade2016,Kolb1990}. To account for the baryon excess, theoretical models must satisfy the Sakharov requirements, one of which is CP symmetry breaking. In the framework of gravitational baryogenesis, CP violation is generated through a coupling that ties the time evolution of the Universe’s geometry to the baryon number current. This framework typically implements an effective interaction in the action written as
\begin{equation}\label{19}
\frac{1}{M_{\star}^{2}}\int d^{4}x\sqrt{-g}J^{\mu}[\partial_{\mu}(R)],
\end{equation}
where $J^{\mu}$ corresponds to the current carrying baryon number, $R$ refers to the Ricci scalar and $M_{\star}$ denotes the high-energy cutoff defining the validity range of the effective model.

Once the matter–antimatter imbalance is generated, its magnitude is characterized by the baryon asymmetry parameter, $\eta_{B}=\eta_{B}-\eta_{\bar{B}}$, which measures the difference between the baryon and antibaryon number densities. In the mechanism of gravitational baryogenesis, it is assumed that the Universe remains in thermal equilibrium while the CP-violating interaction is active. During the cooling of the Universe, when the temperature decreases past the critical decoupling value $T_{D}$, the interactions responsible for baryon production halt, causing the baryon asymmetry to remain constant. Within this scenario, the baryon-to-entropy ratio freezes in with a value given by
\begin{equation}\label{20}
\frac{\eta_{B}}{s}\simeq -\frac{15g_{b}}{4\pi^{2}g_{s}}\bigg(\frac{\dot{R}}{M_{\star}^{2}T_{D}}\bigg),
\end{equation}
In this relation, $g_{s}$ represents the effective number of relativistic degrees of freedom that contribute to the entropy of the thermal plasma, whereas $g_{b}$ specifies the degrees of freedom associated with baryons. The quantity $\dot{R}$ represents the temporal evolution of the Ricci scalar, whereas $M_{\star}$ specifies the cutoff energy scale where CP symmetry–breaking interactions become significant. The resulting expression implies that the baryon-to-entropy ratio is controlled by the temporal evolution of the spacetime curvature during the cosmic expansion. This establishes a deep interplay between the gravitational behavior of the Universe during its early stages and the mechanism that created the present baryon imbalance. The radiation-dominated Universe exhibits a well-defined connection between energy density and temperature:
\begin{equation}\label{21}
\rho(T)=\frac{\pi^{2}}{30}g_{s}T_{D}^{4}.
\end{equation}
\subsection{Baryon asymmetry in $f(T,T_{G})$ gravity}\label{sec3.1}
\hspace{0.5cm} In the $f(T,T_{G})$ formalism, the geometric information of spacetime is captured by the torsion scalar $T$ and the teleparallel Gauss--Bonnet term $T_{G}$ rather than the Ricci scalar. Accordingly, the CP-violating interaction responsible for baryogenesis can be naturally extended by coupling the baryon current to functions of these torsional quantities. A phenomenologically motivated extension of the baryogenesis term may be written as
\begin{equation}\label{22}
\frac{1}{M_{\star}^{2}}\int d^{4}x\sqrt{-g}[\partial_{\mu}(T+T_{G})]J^{\mu}.
\end{equation}
Following the standard equilibrium analysis and assuming that baryon-number-violating interactions freeze out at temperature $T_{D}$, the baryon-to-entropy ratio generated from the above interaction takes the form
\begin{equation}\label{23}
\frac{\eta_{B}}{s}\simeq -\frac{15g_{b}}{4\pi^{2}g_{s}}\bigg(\frac{\dot{T}+\dot{T}_{G}}{M_{\star}^{2}T_{D}}\bigg).
\end{equation}
The values of $\dot{T}$ and $\dot{T}_{G}$ are fixed by the cosmological background solutions of the adopted $f(T,T_{G})$ theory. Consequently, the produced baryon asymmetry becomes directly tied to and reflective of, the modified gravitational dynamics.\\

The minimal torsional description given above captures the basic idea of gravitationally driven baryogenesis, but the $f(T,T_G)$ theory offers a broader perspective where the CP-violating coupling involves the entire modified gravity Lagrangian instead of only the specific scalars $T$ and $T_G$. Under this generalized setup, the interaction term takes the form
\begin{equation}\label{24}
\frac{1}{M_{\star}^{2}}\int d^{4}x\sqrt{-g}[\partial_{\mu}f(T+T_{G})]J^{\mu}.
\end{equation}
The function $f(T,T_{G})$ effectively combines the usual torsion effects represented by $T$ with the more complex, higher-order influences carried by the Gauss–Bonnet–type torsion term $T_G$. When the early Universe remains in thermal balance and baryon-number-changing interactions continue to occur, this mechanism generates an effective baryonic chemical potential, resulting in a nonzero baryon asymmetry.

Assuming freeze-out of these interactions at the decoupling temperature $T_{D}$, the baryon-to-entropy ratio within this generalized picture becomes
\begin{equation}\label{25}
\frac{\eta_{B}}{s}\simeq -\frac{15g_{b}}{4\pi^{2}g_{s}}\bigg(\frac{\dot{T}f_{T}+\dot{T}_{G}f_{T_{G}}}{M_{\star}^{2}T_{D}}\bigg).
\end{equation}
where the derivatives $f_{T}=\frac{\partial f}{\partial T}$ and $f_{T_{G}}=\frac{\partial f}{\partial T_{G}}$ characterize how the gravitational Lagrangian responds to variations in the torsion scalar and the teleparallel Gauss–Bonnet term, respectively.\\

To carry out the baryogenesis study within the $f(T,T_{G})$ gravity framework, one must adopt a specific background evolution for the Universe. In this work, we assume a power-law expansion, characterized by the scale factor
\begin{equation}\label{26}
a(t)=a_{0}t^{m}\Longrightarrow H=\frac{m}{t},
\end{equation}
with $a_{0}>0$ and $m>0$. Choosing a power-law scale factor is cosmologically reasonable because it naturally arises in diverse modified gravity scenarios and can represent several phases of cosmic evolution in a unified manner. In particular, setting $m=\frac{1}{2}$ corresponds to a radiation-dominated Universe, while $m=\frac{2}{3}$ describes the conventional matter-dominated era. For values $m>1$, the scale factor undergoes accelerated expansion, which provides a straightforward model of inflation without the need for an explicit inflation field. The adoption of a power-law form allows us to track the dynamics of $T$ and $T_G$ in a well-controlled and analytically solvable setting, facilitating the evaluation of their time derivatives and the consequent baryon-to-entropy ratio in the context of modified teleparallel gravity.\\

When the power-law expansion is inserted into equation (\ref{25}), the resulting expressions allow us to derive fully analytic forms for the derivatives $f_{T}$, $\dot{f}_{T}$, $f_{T_{G}}$ and $\dot{f}_{T_{G}}$ associated with Model 1.
\begin{equation}\label{27}
f_{T}=\alpha,\hspace{0.3cm} \dot{f}_{T}=0,\hspace{0.3cm} f_{T_{G}}=\frac{\beta t^{2}}{2\sqrt{24m^{3}(m-1)}},\hspace{0.3cm} \dot{f}_{T_{G}}=\frac{\beta t}{\sqrt{24m^{3}(m-1)}}.
\end{equation}
Upon replacing the computed components in equation (\ref{27}) and invoking the Friedmann equation (\ref{15}), the energy density can be written explicitly in analytic form.
\begin{equation}\label{28}
\rho=\frac{3m^{2}}{t^{2}}\bigg[-\alpha+\frac{2m\beta(m+1)}{\sqrt{24m^{3}(m-1)}}\bigg].
\end{equation}
Equating the energy density obtained from our $f(T,T_{G})$ framework (equation (\ref{28})) with the baryogenesis energy density (equation (\ref{21})) leads to a closed-form expression for the decoupling time $t_{D}(T_{D})$.
\begin{equation}\label{29}
t_{D}(T_{D})=\frac{3m}{\pi T_{D}^{2}}\sqrt{\frac{10}{g_{s}}\bigg[-\alpha+\frac{m\beta(m+1)}{\sqrt{6m^{3}(m-1)}}\bigg]}.
\end{equation}
Replacing $t_{D}$ in equation (\ref{23}) with its previously derived expression leads to a closed-form relation for the produced baryon asymmetry within the $f(T,T_{G})$ model.
\begin{equation}\label{30}
\frac{\eta_{B}}{s}\simeq\frac{g_{b}\pi T_{D}^{5}\sqrt{g_{s}}}{6\sqrt{10}mM_{\star}^{2}}\bigg[-\alpha+\frac{m\beta(m+1)}{\sqrt{6m^{3}(m-1)}}\bigg]^{-\frac{3}{2}}\left(1+\frac{4(m-1)\pi^{2}T_{D}^{4}g_{s}}{45m
\bigg[-\alpha+\frac{m\beta(m+1)}{\sqrt{6m^{3}(m-1)}}\bigg]}\right).
\end{equation}
\begin{figure}[h!]
\centering
  \includegraphics[scale=0.5]{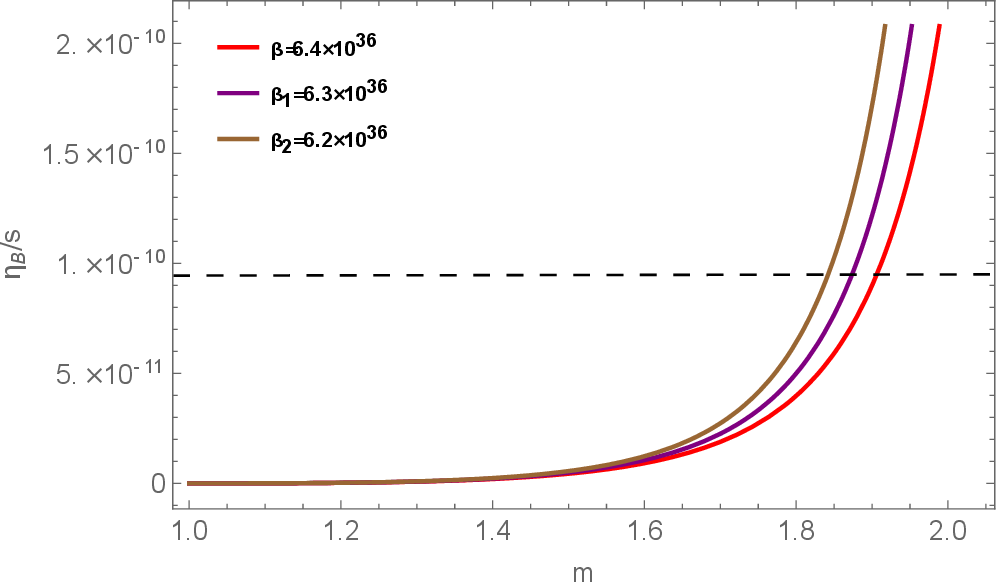}
  \caption{Variation of the baryon-to-entropy ratio $\frac{\eta_{B}}{s}$ with respect to the power-law index $m$ for the \textbf{Model 1} $f(T,T_{G})=\alpha T+\beta \sqrt{T_{G}}$. The curves correspond to three different values of $\beta$, while the other model parameters are fixed at $g_{b}=1$, $g_{s}=106$, $T_{D}=2\times 10^{16}\,\text{GeV}$ and $M_{\star}=2\times 10^{12}\,\text{GeV}$. The plot shows that all parameter choices successfully reproduce the observed constraint $\frac{\eta_{B}}{s}\simeq 9.42\times 10^{-11}$.}
  \label{fig:f1}
\end{figure}

The expression obtained in equation (\ref{30}) is valid under the dynamical assumption that the Universe evolves with a power-law scale factor $a(t)=a_{0} t^{m}$. The derivation contains the combination $\sqrt{6m^{3}(m-1)}$, which requires $m>1$ to keep the solution real and consistent with the assumed background dynamics. Furthermore, the expression remains well-defined only when the model combination $\bigg[-\alpha+\frac{m\beta(m+1)}{\sqrt{6m^{3}(m-1)}}\bigg]$ is positive, ensuring that fractional powers in equation (\ref{30}) do not produce divergences or complex values. These requirements define the admissible parameter space for the baryogenesis solution within $f(T,T_{G})$ gravity. To carry out the numerical analysis, we specify the values of the relevant model parameters $g_{b}=1$, $g_{s}=106$, $T_{D}=2\times 10^{16}\; GeV$, $M_{\star}=2\times 10^{12}\; GeV$ and substitute them into equation (\ref{30}). Using the observational constraint $\frac{\eta_{B}}{s}\simeq 9.42\times 10^{-11}$, we determine the value of the model parameter $\alpha$, obtaining $\alpha\simeq5\times10^{36}$. This value is subsequently used in the numerical analysis of the model.

Next, we explore the dependence of the baryon-to-entropy ratio on the power-law index $m$ by plotting equation (\ref{30}) as a function of $m$ for three different choices of the parameter $\beta$. The resulting curves are displayed in Fig. \ref{fig:f1}. From the plot, it is evident that for all three choices of $\beta$, the predicted baryon asymmetry remains consistent with the observational bound $\frac{\eta_{B}}{s}\simeq 9.42\times 10^{-11}$, which indicates that the model successfully accommodates the observed matter-antimatter asymmetry over a wide range of $m$. The baryon-to-entropy ratio $\frac{\eta_{B}}{s}$ for Model 1 was calculated using three different values of the parameter $\beta$, producing the corresponding scale factor exponents $m = 1.9053, 1.8728, 1.8413$, as listed in Table \ref{Tab:T1}. These three numerical values of $m$ are fully compatible with the theoretical structure of the $f(T,T_{G})$ gravity scenario adopted in this work. Each value lies within the parameter domain where the model remains well-behaved and the background cosmological evolution is mathematically consistent. This confirms that the solutions obtained for Model 1 do not violate the theoretical constraints imposed by the formulation of $f(T,T_{G})$ gravity and therefore represent physically admissible configurations within our framework.
\begin{table}[h!]
\centering
\caption{Numerical values of the model parameters $\alpha$, $\beta$ and $m$ for \textbf{Model 1} producing the observed baryon-to-entropy ratio $\frac{\eta_{B}}{s}\simeq 9.42\times10^{-11}$.}
\begin{tabular}{||p{2.0cm}|p{1.8cm}|p{1.2cm}|p{2.1cm}||}
\hline\hline
\hspace{0.9cm}$\alpha$ & \hspace{0.7cm}$\beta$ & \hspace{0.4cm}$m$ & \hspace{0.8cm}$\frac{\eta_{B}}{s}$\\
\hline\hline
\hspace{0.3cm}$5.0\times10^{36}$ & \hspace{0.1cm}$6.4\times10^{36}$ & $1.9053$ & $9.422\times10^{-11}$\\[1.3pt]
\hline
\hspace{0.3cm}$5.0\times10^{36}$ & \hspace{0.1cm}$6.3\times10^{36}$ & $1.8728$ & $9.422\times10^{-11}$\\[1.3pt]
\hline
\hspace{0.3cm}$5.0\times10^{36}$ & \hspace{0.1cm}$6.2\times10^{36}$ & $1.8413$ & $9.424\times10^{-11}$\\ 
\hline\hline
\end{tabular}
\label{Tab:T1}
\end{table}\\

For generalized baryogenesis in Model 1, using equation (\ref{29}) in equation (\ref{26}) leads to the following expression for $\frac{\eta_{B}}{s}$.
\begin{equation}\label{31}
\frac{\eta_{B}}{s}\simeq\frac{g_{b}\pi T_{D}^{5}\sqrt{g_{s}}}{6\sqrt{10}mM_{\star}^{2}}\bigg[-\alpha+\frac{m\beta(m+1)}{\sqrt{6m^{3}(m-1)}}\bigg]^{-\frac{3}{2}}\left(\alpha+\frac{2m\beta(m-1)}{\sqrt{6m^{3}(m-1)}}\right).
\end{equation}
\begin{figure}[h!]
\centering
  \includegraphics[scale=0.5]{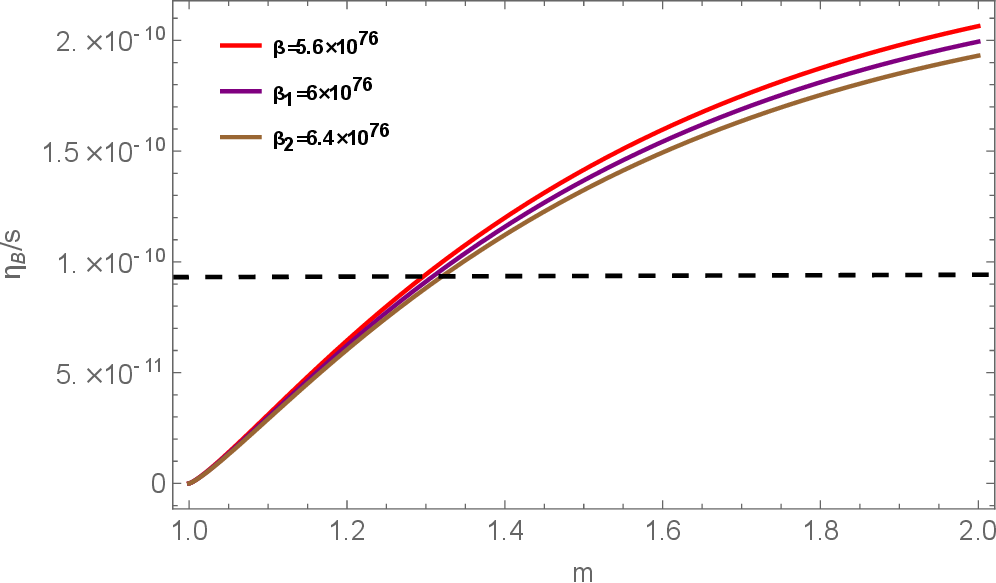}
  \caption{Plot of the baryon-to-entropy ratio $\frac{\eta_{B}}{s}$ versus the power–law parameter $m$ for the generalized baryogenesis scenario of \textbf{Model 1} $f(T,T_{G})=\alpha T+\beta \sqrt{T_{G}}$. All curves satisfy the observational bound $\frac{\eta_{B}}{s}\simeq 9.42\times 10^{-11}$.}
  \label{fig:f2}
\end{figure}

We note that the generalized expression inherits the same physical admissibility conditions derived earlier for equation (\ref{30}). The requirement $m>1$ guarantees a real power-law background, while the positivity of the term inside the fractional power prevents divergences or complex values. Thus, no new constraints arise beyond those previously established. For uniformity across the analysis, we evaluate the generalized baryogenesis relation with the same parameters used in the earlier section, incorporating the earlier derived value of $\alpha$. This allows a coherent comparison of results throughout the $f(T,T_{G})$ model. The subsequent step is to study the response of the baryon-to-entropy ratio $\frac{\eta_{B}}{s}$ to changes in the power-law parameter $m$ within the context of the generalized baryogenesis mechanism in Model 1. Figure \ref{fig:f2} presents the behavior of the baryon-to-entropy ratio $\frac{\eta_{B}}{s}$ as a function of $m$, evaluated for three representative coupling strengths, $\beta = 5.6 \times 10^{76}$, $6.0 \times 10^{76}$ and $6.4 \times 10^{76}$, while keeping $\alpha$ fixed at its previously established value. Inspection of the figure reveals that all three trajectories satisfy the empirical bound $\frac{\eta_{B}}{s}\simeq 9.42 \times 10^{-11}$. This agreement affirms the compatibility of the generalized baryogenesis formulation with the underlying $f(T,T_{G})$ gravitational theory. As reported in Table \ref{Tab:T2}, the generalized framework gives rise to $m$ values of $1.2996$, $1.3117$ and $1.3235$ corresponding to the three choices of $\beta$. It is also important to note that the computed values of $m$ with respect to the theoretical restriction $m>1$. This condition prevents the term $\sqrt{6m^{3}(m-1)}$ from becoming non-real and guarantees the coherence of the generalized baryogenesis formulas. Hence, the obtained solutions are compatible not only with observational data but also with the intrinsic consistency requirements of the $f(T,T_{G})$ gravitational framework.
\begin{table}[h!]
\centering
\caption{Numerical values of the model parameters $\alpha$, $\beta$ and $m$ and the resulting baryon-to-entropy ratio $\frac{\eta_{B}}{s}$ for the generalized baryogenesis case of \textbf{Model 1}.}
\begin{tabular}{||p{2.0cm}|p{1.8cm}|p{1.2cm}|p{2.1cm}||}
\hline\hline
\hspace{0.9cm}$\alpha$ & \hspace{0.7cm}$\beta$ & \hspace{0.4cm}$m$ & \hspace{0.8cm}$\frac{\eta_{B}}{s}$\\
\hline\hline
\hspace{0.3cm}$5.0\times10^{36}$ & \hspace{0.1cm}$5.6\times10^{76}$ & $1.9053$ & $9.420\times10^{-11}$\\[1.3pt]
\hline
\hspace{0.3cm}$5.0\times10^{36}$ & \hspace{0.1cm}$6.0\times10^{76}$ & $1.8728$ & $9.422\times10^{-11}$\\[1.3pt]
\hline
\hspace{0.3cm}$5.0\times10^{36}$ & \hspace{0.1cm}$6.4\times10^{76}$ & $1.8413$ & $9.420\times10^{-11}$\\ 
\hline\hline
\end{tabular}
\label{Tab:T2}
\end{table}\\

Proceeding to Model 2, substitution of the functional form from equation (\ref{25}) enables the computation of closed-form analytical expressions for the quantities $f_{T}$, $\dot{f}_{T}$, $f_{T_{G}}$ and $\dot{f}_{T_{G}}$.
\begin{equation}\label{32}
f_{T}=-1,\hspace{0.3cm} \dot{f}_{T}=0,\hspace{0.3cm} f_{T_{G}}=\delta\log\left(\frac{24m^{3}(m-1)}{t^{4}}\right),\hspace{0.3cm} \dot{f}_{T_{G}}=-\frac{4\delta}{t}.
\end{equation}
Using the derivatives evaluated above and substituting them into equation (\ref{15}), we derive the explicit form of the energy density for Model 2.
\begin{equation}\label{33}
\rho=\frac{3m^{2}}{t^{2}}-\frac{48\delta m^{3}}{t^{4}}.
\end{equation}
The decoupling time can be determined by matching the Model 2 energy density in equation (\ref{33}) to the baryogenesis energy density described in equation (\ref{21}). Solving this equality provides a closed-form expression for $t_{D}$ in terms of the decoupling temperature $T_{D}$.
\begin{equation}\label{34}
t_{D}=\frac{\sqrt{3}m}{\pi T_{D}^{2}\sqrt{g_{s}}}\sqrt{\left[15+\sqrt{\frac{5}{m}}\sqrt{45m+32\pi^{2}g_{s}T_{D}^{4}\delta}\right]}.
\end{equation}
On substituting $t_{D}$ into equation (\ref{23}), the expression for $\frac{\eta_{B}}{s}$ becomes:
\begin{equation}\label{35}
\frac{\eta_{B}}{s}\simeq\frac{5\sqrt{3}g_{b}\pi T_{D}^{5}\sqrt{g_{s}}}{mM_{\star}^{2}}\left[15+\sqrt{\frac{5}{m}}\sqrt{45m+32\pi^{2}g_{s}T_{D}^{4}\delta}\right]^{-\frac{3}{2}}
\left(1+\frac{8(m-1)\pi^{2}g_{s}T_{D}^{4}}{3m\left[15+\sqrt{\frac{5}{m}}\sqrt{45m+32\pi^{2}g_{s}T_{D}^{4}\delta}\right]}\right).
\end{equation}
\begin{figure}[h!]
\centering
  \includegraphics[scale=0.5]{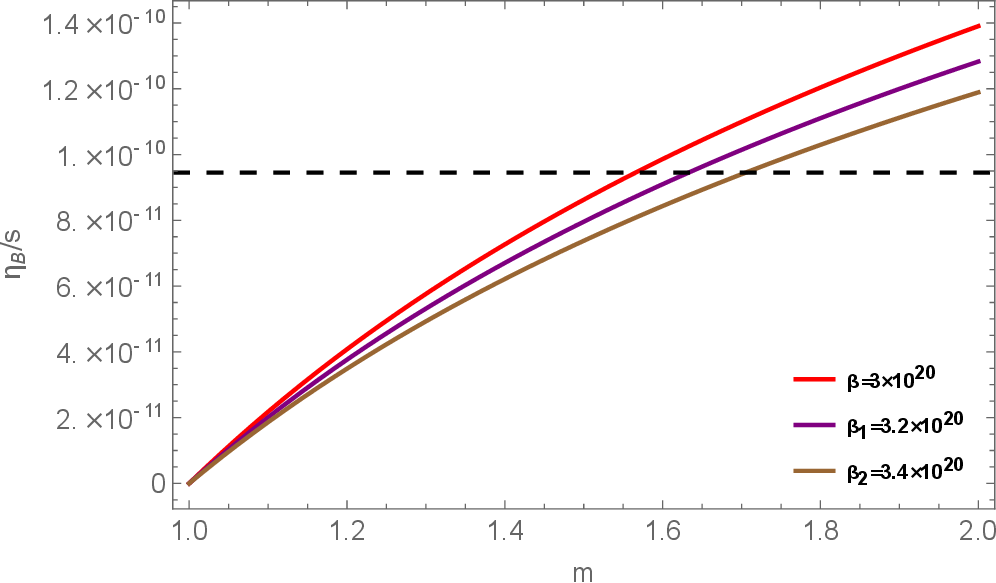}
  \caption{Plot of the baryon-to-entropy ratio $\frac{\eta_{B}}{s}$ versus the power–law parameter $m$ for the \textbf{Model 2} $f(T,T_{G})=-T+\delta\; T_{G}\;\log(T_{G})$. All trajectories intersect the observational bound $\frac{\eta_{B}}{s}\simeq 9.42\times 10^{-11}$, demonstrating the viability of the model across the chosen parameter space.}
  \label{fig:f3}
\end{figure}

The baryon-to-entropy ratio obtained in equation (\ref{35}) is well defined only within the parameter domain consistent with the assumed power-law background $a(t)=a_{0} t^{m}$. To maintain mathematical consistency, the model imposes the constraint $m>1$, which guarantees that all torsional contributions remain real and that the formulas involving square-root or fractional terms do not develop singularities. Additionally, ensuring that $\left[15+\sqrt{\tfrac{5}{m}}\sqrt{45m+32\pi^{2}g_{s}T_{D}^{4}\delta}\right] > 0$ keeps the expression finite and allows equation (\ref{35}) to produce valid values of $\frac{\eta_{B}}{s}$. Adopting the parameter choices $g_{b}=1$, $g_{s}=106$, $T_{D}=2\times10^{16}\; \mathrm{GeV}$ and $M_{\star}=2\times10^{12}\; \mathrm{GeV}$ and enforcing the observational requirement $\frac{\eta_{B}}{s}\simeq 9.42\times10^{-11}$, we find that the parameter $\delta$ must satisfy the bound $\delta>2.89985\times10^{20}$. For a deeper analysis, the baryon-to-entropy ratio $\frac{\eta_{B}}{s}$ is plotted as a function of $m$ for three selected values of the parameter $\delta$. From Figure \ref{fig:f3}, it is evident that $\delta = 3.0 \times 10^{20}$, $3.2 \times 10^{20}$ and $3.4 \times 10^{20}$ produce results compatible with the observed baryogenesis constraint, associated with $m$ values $1.5632$, $1.6300$ and $1.7020$, respectively. The outcomes are compiled in Table \ref{Tab:T3}. All obtained $m$ values meet the requirement $m>1$, ensuring the background cosmology remains consistent and showing that Model–II within $f(T,T_{G})$ gravity reproduces the observed baryon-to-entropy ratio.
\begin{table}[h!]
\centering
\caption{Numerical values of $\delta$, $m$ and the resulting baryon-to-entropy ratio $\frac{\eta_{B}}{s}$ for \textbf{Model 2}.}
\begin{tabular}{||p{2.0cm}|p{1.2cm}|p{2.1cm}||}
\hline\hline
\hspace{0.9cm}$\delta$ & \hspace{0.4cm}$m$ & \hspace{0.8cm}$\frac{\eta_{B}}{s}$\\
\hline\hline
\hspace{0.3cm}$3.0\times10^{20}$  & $1.5632$ & $9.420\times10^{-11}$\\[1.3pt]
\hline
\hspace{0.3cm}$3.2\times10^{20}$  & $1.63$ & $9.420\times10^{-11}$\\[1.3pt]
\hline
\hspace{0.3cm}$3.4\times10^{20}$ & $1.702$ & $9.421\times10^{-11}$\\ 
\hline\hline
\end{tabular}
\label{Tab:T3}
\end{table}\\

For the generalized Model 2 baryogenesis, the baryon-to-entropy ratio $\frac{\eta_{B}}{s}$ is determined by plugging the modified decoupling time from equation (\ref{34}) into equation (\ref{26}).
\begin{eqnarray}\label{36}
\frac{\eta_{B}}{s}&\simeq&-\frac{5\sqrt{3}g_{b}\pi T_{D}^{5}\sqrt{g_{s}}}{mM_{\star}^{2}}\left[15+\sqrt{\frac{5}{m}}\sqrt{45m+32\pi^{2}g_{s}T_{D}^{4}\delta}\right]^{-\frac{3}{2}}\\\nonumber
&&\left(1-\frac{8(m-1)\delta\pi^{2}g_{s}T_{D}^{4}}{3m\left[15+\sqrt{\frac{5}{m}}\sqrt{45m+32\pi^{2}g_{s}T_{D}^{4}\delta}\right]}\times
\log\left[\frac{8(m-1)\pi^{4}g_{s}^{2}T_{D}^{8}}{3m\left[15+\sqrt{\frac{5}{m}}\sqrt{45m+32\pi^{2}g_{s}T_{D}^{4}\delta}\right]^{2}}\right]\right).
\end{eqnarray}
\begin{figure}[h!]
\centering
  \includegraphics[scale=0.5]{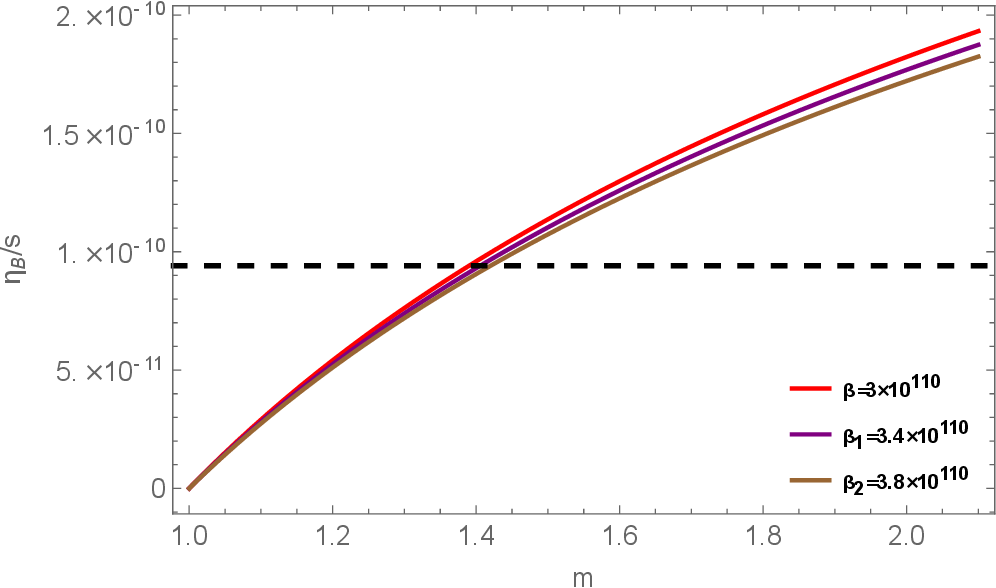}
  \caption{Plot of the baryon-to-entropy ratio $\frac{\eta_{B}}{s}$ as a function of the model parameter $m$ of the \textbf{Model 2} $f(T,T_{G})=-T+\delta\; T_{G}\;\log(T_{G})$ for three different values of $\delta$. The plot shows that observational baryogenesis criteria can be satisfied within well-defined ranges of $m$ for each choice of $\delta$.}
  \label{fig:f4}
\end{figure}

To further explore the predictions of Model 2 within generalized baryogenesis, Figure \ref{fig:f4} presents the baryon-to-entropy ratio $\frac{\eta_{B}}{s}$ as a function of $m$, considering multiple values of the parameter $\delta$. The plot, based on our modified gravitational expression for $\frac{\eta_{B}}{s}$, demonstrates how the baryon asymmetry responds to variations in the model’s admissible parameter space. The analysis considers three sample values for the coupling $\delta$, specifically $3\times10^{110}$, $3.4\times10^{110}$ and $3.8\times10^{110}$. For each of these choices, the parameter $m$ is varied continuously until the predicted value of $\frac{\eta_{B}}{s}$ satisfies the observational baryogenesis bound $\frac{\eta_{B}}{s}\simeq 9 \times 10^{-11}$. The resulting values of $m$ that satisfy this constraint are found to be $m = 1.3908$, $1.4064$ and $1.4208$ for the three respective $\delta$ values. The complete set of results is summarized in Table \ref{Tab:T4}, which lists the values of $\delta$, the corresponding allowed $m$ values, and the resulting baryon-to-entropy ratio $\frac{\eta_{B}}{s}$. From these results, it is evident that all obtained values of $m$ remain above unity, satisfying the theoretical constraint $m>1$ required by the structure of the model. Therefore, the analysis demonstrates that Model 2 successfully reproduces the observed baryon asymmetry of the Universe within realistic and mathematically consistent ranges of the free parameters.
\begin{table}[h!]
\centering
\caption{Numerical values of the coupling parameter $\delta$, the corresponding allowed values of the model parameter $m$ and the resulting baryon-to-entropy ratio $\frac{\eta_{B}}{s}$ for \textbf{Model 2}.}
\begin{tabular}{||p{2.0cm}|p{1.2cm}|p{2.1cm}||}
\hline\hline
\hspace{0.9cm}$\delta$ & \hspace{0.4cm}$m$ & \hspace{0.8cm}$\frac{\eta_{B}}{s}$\\
\hline\hline
\hspace{0.3cm}$3.0\times10^{110}$  & $1.3908$ & $9.421\times10^{-11}$\\[1.3pt]
\hline
\hspace{0.3cm}$3.4\times10^{110}$  & $1.4064$ & $9.421\times10^{-11}$\\[1.3pt]
\hline
\hspace{0.3cm}$3.8\times10^{110}$ & $1.4208$ & $9.421\times10^{-11}$\\ 
\hline\hline
\end{tabular}
\label{Tab:T4}
\end{table}
\section{Final remarks and physical implications}\label{sec4}
\hspace{0.5cm} The present investigation explored the viability of baryogenesis and generalized baryogenesis within the framework of $f(T,T_{G})$ gravity by examining two representative modified gravity models, $f(T,T_{G})=\alpha T+\beta\sqrt{T_{G}}$ (Model 1) and $f(T,T_{G})=-T+\delta\; T_{G}\;\log(T_{G})$ (Model 2). Under the assumption of a power-law expansion $a(t) = a_0 t^m$, the modified field equations were solved to obtain explicit expressions for the baryon-to-entropy ratio $\eta_{B}/s$. Using the standard high-energy choices $g_b = 1$, $g_s = 106$, $T_D = 2 \times 10^{16}$ GeV and $M_\star = 2 \times 10^{12}$ GeV, we systematically examined whether the models can reproduce the observational requirement $\eta_{B}/s \simeq 9.42 \times 10^{-11}$. 

For Model 1, the standard baryogenesis mechanism yields an analytically well-behaved expression only when $m>1$ and the parameter combination inside the fractional powers remains positive, preventing singular or non-real solutions. After imposing the observational constraint, we obtained the value $\alpha \simeq 5 \times 10^{36}$, which sets the appropriate torsional contribution during the decoupling phase. A detailed numerical exploration, supported by Figure \ref{fig:f1} and Table \ref{Tab:T1} included in the study, revealed that the baryogenesis relation is satisfied for multiple choices of $\beta$, giving rise to viable values of $m$ such as $1.9053$, $1.8728$ and $1.8413$. Since all these values lie in the theoretically allowed region $m>1$, Model 1 not only reproduces the observed baryon asymmetry but also remains fully consistent with the internal mathematical structure of $f(T,T_{G})$ gravity.

A similar conclusion emerges from the generalized baryogenesis scenario of Model 1. The generalized expression inherits the same physical constraints without introducing additional restrictions. For the same set of model parameters, the baryon-to-entropy ratio in this extended framework again matches observations for suitable parameter combinations such as $\beta = 5.6 \times 10^{76}$, $6.0 \times 10^{76}$ and $6.4 \times 10^{76}$, corresponding to values of $m = 1.2996$, $1.3117$, and $1.3235$. These values again exceed unity, which confirm that the model supports a mathematically consistent cosmic background while simultaneously satisfying current observational limits.

Model 2, based on the functional form $f(T,T_{G})=-T+\delta\; T_{G}\;\log(T_{G})$, exhibits a similar behavior. The mathematical structure again requires $m>1$ to ensure that square-root and power-law terms remain real. By imposing the observational baryogenesis condition, we derived a lower bound $\delta > 2.89985 \times 10^{20}$. Subsequent numerical analysis showed that values such as $\delta = 3.0 \times 10^{20}$, $3.2 \times 10^{20}$ and $3.4 \times 10^{20}$ yield baryon-to-entropy ratios consistent with observation, corresponding to allowed values $m = 1.5632$, $1.6300$ and $1.7020$. The same conclusion holds for generalized baryogenesis in Model~II, where larger allowed values of $\delta$ lead to physically admissible $m$-ranges such as $1.3908$, $1.4064$ and $1.4208$. In all cases, the admissible solutions satisfy $m>1$, ensuring that the cosmological background remains theoretically consistent.

A central result of this work is that both gravitational frameworks can successfully account for the observed baryon asymmetry without invoking exotic matter components, additional scalar fields, or external CP-violating interactions beyond those naturally introduced by torsion. The analysis demonstrates that the modifications of general relativity encoded in $f(T,T_G)$ gravity inherently provide the mathematical structure required to produce a non-zero baryon-to-entropy ratio at the observed scale. Moreover, both models feature sufficiently broad parameter spaces, allowing for viable solutions without the need for extreme fine-tuning, thereby reinforcing the physical plausibility of torsion-based extensions as alternatives to conventional baryogenesis mechanisms. From a wider perspective, the concordance between theoretical predictions and observational data highlights the significant role of torsion and the teleparallel Gauss–Bonnet term in shaping the physics of the early Universe, particularly during epochs when standard curvature-based formulations may be insufficient to capture all gravitational degrees of freedom. The power-law cosmological background employed in this study provides a simple yet effective framework to describe early-time cosmic expansion.

In summary, the present analysis demonstrates that both proposed forms of $f(T,T_G)$ gravity successfully reproduce the observed baryon-to-entropy ratio within realistic and theoretically consistent parameter domains. This reinforces the viability of torsion-based modified gravity as a promising framework for addressing the long-standing matter-antimatter asymmetry problem. Future research may incorporate perturbation analyses, non-power-law expansions, reheating mechanisms, or quantum-gravity corrections to further test the predictive scope of these models and to develop a more comprehensive understanding of the role of torsion in the early Universe.

\end{document}